What is Overlap Weighting, How Has it Evolved, and When to Use It for Causal Inference?

**Authors:** Haidong Lu[1,2], Fan Li[3], Laine E. Thomas[4,5], Fan Li[6,7]

**Affiliations:**

[1] Department of Internal Medicine, Yale School of Medicine, New Haven, CT, USA

[2] Department of Chronic Disease Epidemiology, Yale School of Public Health, New Haven, CT, USA

[3] Department of Statistical Science, Duke University, Durham, NC, USA

[4] Department of Biostatistics and Bioinformatics, Duke University, Durham, NC, USA

[5] Duke Clinical Research Institute, Durham, NC, USA

[6] Department of Biostatistics, Yale School of Public Health, New Haven, CT, USA

[7] Center for Methods in Implementation and Prevention Science, Yale School of Public Health, New Haven, CT, USA

**Correspondence**: Haidong Lu, Department of Internal Medicine, Yale School of Medicine, 367 Cedar Street, New Haven, CT 06510 (email: haidong.lu@yale.edu); Fan Li, Department of Biostatistics, Yale School of Public Health, Suite 200, Room 229, New Haven, CT 06510 (email: fan.f.li@yale.edu)

**Data availability**: No data were generated or analyzed in support of this research.

**Author contributions**: H.L. and F.L. (Yale) contributed to the conception and design of the study, funding acquisition, and the original draft of the manuscript. F.L. (Duke) and L.E.T. contributed to critical review of the manuscript.

**Use of Artificial Intelligence (AI) tools**: In the preparation of this manuscript, the AI tool GPT-5 was employed to enhance the readability of specific sentences.

**Funding**: This work was supported by grants from National Institutes of Health [grant numbers R00DA057487 to H.L. and R01HL168202 to F.L. (Yale)].

**Disclaimer:** All statements in this report, including its findings and conclusions, are solely those of the authors and do not necessarily represent the official views of the National Institutes of Health.

**Conflict of interest**: None declared.

**Summary**

The growing availability of large health databases has expanded the use of observational studies for comparative effectiveness research. Unlike randomized trials, observational studies must adjust for systematic differences in patient characteristics between treatment groups. Propensity score methods, including matching, weighting, stratification, and regression adjustment, address this issue by creating groups that are comparable with respect to measured covariates. Among these approaches, overlap weighting (OW) has emerged as a principled and efficient method that emphasizes individuals at empirical equipoise, those who could plausibly receive either treatment. By assigning weights proportional to the probability of receiving the opposite treatment, OW targets the Average Treatment Effect in the Overlap population (ATO), achieves exact mean covariate balance under logistic propensity score models, and minimizes asymptotic variance. Over the last decade, the OW method has been recognized as a valuable confounding adjustment tool across the statistical, epidemiologic, and clinical research communities, and is increasingly applied in clinical and health studies. Given the growing interest in using observational data to emulate randomized trials and the capacity of OW to prioritize populations at clinical equipoise while achieving covariate balance (fundamental attributes of randomized studies), this article provides a concise overview of recent methodological developments in OW and practical guidance on when it represents a suitable choice for causal inference.

**Keywords**: overlap weight; propensity score; observational study; causal inference; epidemiologic research

**Key Messages**

- **What is the method about?**

    Overlap weighting (OW) is a propensity score-based method that assigns weights

proportional to the probability of receiving the opposite treatment, emphasizing individuals who could plausibly receive either treatment and estimating the Average Treatment Effect in the Overlap population (ATO) with exact mean covariate balance and optimal efficiency.

- **Why is it important?**

  OW is important because it provides a principled and interpretable solution to the problem of limited covariate overlap in observational studies, reducing instability from extreme propensity scores, improving precision compared with inverse probability weighting, and aligning results with meaningful populations. Recent developments in methodological research and applied work have extended OW to multiple treatments, time-to-event outcomes, causal subgroup analysis, and so forth.

- **Where or how is it best used (or not used)?**

  OW is best used in observational studies with partial but nonzero overlap, where the goal is to estimate treatment effects among individuals at clinical equipoise, because it provides stable, statistically efficient, and interpretable estimates by emphasizing comparable patients and avoiding arbitrary data trimming.

The growing availability of large clinical databases has stimulated broader use of observational treatment comparisons. Unlike randomized controlled trials which assure that patient characteristics are similar across treatment arms, observational studies must attempt to adjust for bias due to differences in characteristics between treated and untreated patients. Methods that adjust for these differences include multivariable regression modeling and propensity score methods. The propensity score is a summary of factors associated with receiving treatment that can be used for matching, weighting, stratification, and regression adjustment.[1–4] Propensity score methods have been advocated because they separate the design phase, where treatment groups are made comparable, from the analysis of outcomes, hence ensuring objectivitiy.[5]

Whether to use matching, weighting, stratification, or regression adjustment on the propensity score, and which variation to use, is often subjective in practice. Although these approaches can produce qualitatively similar results, the decision regarding the primary analytic strategy can be consequential when a different point estimate or precision leads to different study conclusions.[5,6] In this education corner article, we focus on demystifying a recent propensity score weighting method—the overlap weighting (OW). Among the class of propensity score methods, OW optimizes attributes related to bias and precision. The OW method has been introduced as a useful confounding adjustment tool among the statistics, and epidemiology and clinical research communities in the last decade[6–8], and has been increasingly used in clinical and health studies. Given the increasing popularity of conducting observational studies to emulate randomized trials[9], and the capacity of OW to target the population at clinical equipoise in addition to achieving covariate balance (both of which are fundamental attributes of randomized trials), we aim to provide a one-stop shop on its recent advances and explain when OW is a good choice for causal inference.

**DEFINITION AND INTUITION OF OVERLAP WEIGHT VERSUS INVERSE PROBABILITY OF TREATMENT WEIGHTING**

Inverse probability of treatment weighting (IPTW) creates a pseudo-population in which treatment assignment is independent of measured baseline covariates. In IPTW, each individual is weighted by the inverse of the probability of receiving the treatment they actually received, conditional on their covariates. Assuming that $X$ is a vector of observed pretreatment covariates, which suffices to control for all sources of confounding. The propensity score, $e(X) = P(Z = 1|X)$, is the probability of receiving the treatment $Z = 1$ conditional on the baseline covariates. In a binary treatment setting, the weight is $1/e(X)$ for treated individuals and $1/(1 - e(X))$ for untreated individuals. By reweighting the sample in this way, IPTW estimates the average treatment effect (ATE) in the target population represented by the combined study sample. This weighting scheme "up-weights" individuals who are underrepresented in their treatment group relative to what would be expected under randomization, and "down-weights" those who are overrepresented.

IPTW can become numerically unstable when the estimated propensity scores are very close to 0 or 1, leading to extremely large weights, even when the stabilized version (i.e., incorporating the marginal probability of receiving the treatment actually received in the numerator) is employed. Such extreme weights often arise when there is limited overlap in the covariate distributions between treatment groups, a phenomenon also known as a practical violation of the positivity assumption.[10,11] In these situations, IPTW estimates often have high variance and become sensitive to model misspecification, as this method relies heavily on individuals in the tails of the propensity score distribution.

In contrast, OW addresses this numerical instability issue from a design perspective, by assigning weights proportional to the probability of receiving the opposite treatment, rather than the inverse probability of receiving the actual treatment. Specifically, for treated individuals, the overlap weight is $1 - e(X)$, and for untreated individuals, it is $e(X)$; that is the weight is the propensity to receive the other treatment. In conceiving the target population, this approach effectively emphasizes individuals in the region of covariate overlap between the treatment groups (i.e. those whose propensity scores are near 0.5) and de-emphasize individuals whose covariate profiles make them almost certain to receive one treatment or the other (i.e., lying on the tails of the propensity score distribution). As a result, OW estimates the average treatment effect in the overlap population (ATO), which is defined as the subpopulation with the greatest covariate overlap, mimicking a target population with equipoise for receiving treatment.

The intuition behind OW is that when our goal is to estimate causal effects to inform treatment decisions, the most informative subjects are those who could plausibly have received either treatment. These are the individuals who are most comparable across treatment arms, and thus least likely to require heavy extrapolation based on the propensity score model. Conceptually, patients for whom the optimal treatment is most uncertain—those "on the fence" between alternatives—are precisely the ones for whom comparative effectiveness evidence is most informative. Although clinical equipoise, or uncertainty regarding the superior treatment, is a prerequisite for randomized trial participation,[12] patients with propensity scores close to 0 or 1 empirically violate this principle by exhibiting little to no treatment uncertainty, and are therefore less emphasized by OW. Statistically, the available information *on patients with extreme propensity scores* is largely confined to a single treatment arm, with almost no comparator data. By down-weighting individuals with extreme propensity scores near 0 or 1, OW has been mathematically proved (and empirically demonstrated) to yield a causal effect estimator with the smallest variance and greater efficiency, while achieving exact balance in the means of all measured covariates between the weighted treatment groups.[6,7]

In summary, IPTW targets the ATE and incorporates all individuals in the pseudo-population, but it can be sensitive to poor overlap and extreme weights. OW, by contrast, targets the ATO, prioritizes comparability and internal validity, and is more efficient when treatment groups differ substantially in baseline characteristics and have limited overlap. The discrepancy between the ATE from IPTW and the ATO from OW depends on two key factors: the degree of overlap and the extent of treatment effect heterogeneity. In an ideal randomized controlled trial with 1:1 allocation, perfect overlap at baseline ensures that ATE and ATO are identical, regardless of treatment effect heterogeneity. In contrast, when there is some degree of non-overlap (as illustrated in Figure 1), ATE and ATO can diverge, especially in the presence of strong treatment effect heterogeneity. Importantly, while OW and IPTW provide complementary perspectives on treatment effects, examining the discrepancies between ATE and ATO can yield additional insights. In particular, large discrepancies between ATE and ATO estimates should prompt careful reflection on the degree of non-overlap and on which notion of treatment benefit is most meaningful given the context.

In the following section, we will provide a one-stop overview of the existing developments of OW, and discuss how it has elvoved (see Table 1 as a brief summary).

**EXISTING DEVELOPMENTS OF OW IN THE LITERATURE**

**Origins of OW in Statistics**

While the issue of limited overlap and the OW had been studied and acknowledged in earlier work[13–18], the concept of OW and ATO was formally introduced by Li et al. in 2018.[6] Specifically, Li et al.

defined and developed the balancing weights framework, which unifies various propensity score weighting methods under a common structure. Within this framework, OW emerge as a member that leads to the weighting estimator with the smallest asymptotic variance and highest asymptotic efficiency. For a binart treatment, when the propensity score is modeled by logistic regression, OW also yields exact mean balance of covariates between treatment groups. OW has been found to provide a clinically interpretable estimand, that is, the treatment effect among patients who could plausibly receive either treatment. This conceptual innovation has since motivated a growing body of methodological research and applied work that extends OW to multiple treatments, time-to-event outcomes, clustered data, and high-dimensional covariates.

**Simulation Evidence**

The first systematic evaluation of overlap weights (OW) was presented by Li et al.[7] This paper conducted simulation studies to compare OW against alternative methods for handling limited overlap, including IPTW with and without trimming at the tails of propensity score distribution. The results demonstrated that OW consistently achieved superior covariate balance, reduced variance, and more stable causal effect estimates, particularly in settings with extreme propensity scores. Another contribution of this work was the derivation of a closed-form variance estimator for OW when the propensity score is estimated using logistic regression. This development allows for straightforward inference without relying solely on computationally intensive resampling methods.

While Li et al. provided a closed-form variance estimator for OW under logistic regression propensity scores[7], subsequent work has broadened the discussion. Austin compared bootstrap and asymptotic variance estimation when using propensity score weighting, including weights for estimating ATE and the average treatment effect in the treated (ATT), matching weights, and OW, for both continuous and binary outcomes.[19] The findings underscore that for OW, the closed-form symptotic variance estimator can perform comparably to the bootstrap even when sample size is small. In parallel, Austin addressed the issue of power and sample size calculations for propensity score weighting, including overlap weights, with regards to within-person homogeneity in outcomes.[20] Specifically, the authors provided a method that estimates the variance inflation factor and accounts for the reduction in effective sample size caused by weighting. Liu et al. recently provided a more general framework for power calculation under propensity score weighting, including OW.[21] These empirical evidence complement the original work of Li et al. by making OW more accessible and implementable in real-world applications.

**OW with Multiple Treatments**

As many epidemiologic studies involve comparisons across more than two treatment options, Li and Li extended the balancing weights framework to the multiple treatment setting (including factorial assignments as special cases). They introduced the generalized overlap weights that emphasize individuals most likely to receive any of the available treatments.[22] The generalized overlap weights are formulated as the original IPTW multiplied by the harmonic mean of the generalized propensity scores[23]. The generalized OW have also been identified as the member of the balancing weights that minimize the total variance for estimating all pairwise causal effects contrasts. The generalized ATO is then intepreted as the estimand among the subpopulation with substantial overlap to receive all treatments. Building on this work, Yu et al. conducted a systematic comparison of parametric IPTW, matching weights, and generalized OW, with multiple treatments.[24]

**OW with Time-to-event Outcomes**

Time-to-event outcomes present unique challenges for propensity score weighting, as censoring must be accounted for when estimating counterfactual survival functions. Early work by Mao et al. provided a unified analytic framework for propensity score weighting with time-to-event outcomes, advocating and clarifying target estimands and proposing inference procedures.[25] This work incorporated the class of balancing weights (including OW) into the definition and estimation of causal estimands, and demonstrated that combining OW with a Cox outcome model can yield substantial efficiency gains over IPTW across a range of causal estimands. It laid important groundwork for incorporating overlap weights into survival analysis. Relatedly, marginal Cox models estimated via inverse probability weighting have been studied and applied for time-to-event outcomes, for example by Austin and Stuart[26]. The extension of this framework to OW is conceptually straightforward, and directly targets marginal causal hazard ratio in the region of covariate overlap (with the caveat that the hazard ratio is causally interpretable under the proportional hazards assumption[27,28]). Such OW-weighted marginal Cox models have already been adopted in many applied survival analyses, including recent implementations available in the *PSsurvival* R package.[29]

In the context of covariate-dependent censoring, Cheng et al. proposed estimators that combine inverse probability of censoring weighting (IPCW) with propensity score weighting, including IPTW with and without trimming and OW, to estimate counterfactual survival functions.[30] Through simulations, they showed that OW is the most efficient and consistently outperforms IPTW with and without trimming. Cao et al. combined IPCW with overlap weights to estimate restricted mean survival times, a clinically interpretable summary measure of treatment effect.[31] Zeng et al. introduced a pseudo-observation approach under propensity score weighting, including OW, and developed the large-sample theory of the final estimators.[32] Together, these contributions establish a growing toolkit for applying overlap weights to survival analysis.

**Combing OW with Outcome Regression**

While OW achieve exact covariate balance in means under a correctly specified propensity score model, efficiency can be further improved by incorporating information from the outcome model. Mao et al. proposed augmented OW, which combine outcome regression with OW in a doubly robust framework.[33] This approach yields consistent estimation if either the propensity score model or the outcome regression model is correctly specified, and it typically improves efficiency compared to OW alone. Zeng et al. further demonstrated the utility of augmented OW in survival analysis using pseudo-observations, showing that the combination of OW and outcome regression can enhance precision and robustness in estimating causal effects under censoring in survival contexts.[32]

Relatedly, the connection between overlap weighting and regression adjustment for the propensity score has been clarified in both classical and recent work. Vansteelandt and Daniel[17] demonstrated that, in linear models adjusting for the estimated propensity score, the coefficient for the treatment corresponds to a *weighted* average treatment effect, emphasizing that weighting by $e(x)\{1 - e(x)\}$ (the "optimal weight") yields a consistent and interpretable estimand even when the outcome model is misspecified. This optimal weight defines the same target population as ATO, which OW explicitly estimates. More recently, Ding[34] showed that in a simple regression that adjusts for propensity score as a covariate, the coefficient on treatment also identifies the ATO under large-sample conditions even if the outcome model is incorrect, providing a unifying view of OW and propensity-score-adjusted regression.

**OW for Randomized Clinical Trials**

Although propensity score weighting methods are most commonly applied to observational studies, they can also play a valuable role in randomized clinical trials by improving statistical efficiency through covariate adjustment. Zeng et al. examined the use of propensity score weighting for covariate adjustment in RCTs, including OW and IPTW.[35] First, as a central observation, OW and IPTW target the same ATE estimand under randomization. Second, although randomization balances covariates in expectation, chance imbalances in baseline covariates can occur, particularly in small to moderate sample sizes. Incorporating OW into the analysis framework provides an alternative way to adjust for baseline covariates and to improve precision. Although OW and IPTW estimators are asymptotically equivalent in large samples, surprisingly, OW leads to more efficient estimates of treatment effects compared to IPTW estimator in finite samples. This approach has been recently extended to subgroup analysis[36] and win statistics in randomized clinical trials[37], and to cluster randomized trials[38].

**OW in Causal Subgroup Analysis**

Causal subgroup analysis seeks to evaluate treatment effect heterogeneity across predefined or data-driven subgroups, a topic of increasing importance in precision medicine. Yang and colleagues have extended propensity score weighting, including overlap weights (OW), to this setting.[39,40,36] They developed a suite for causal subgroup analysis with binary outcomes, and introduced the subgroup weighted average treatment effect (S-WATE) as a formal estimand.[39] They developed the OW-pLASSO algorithm, which combines overlap weighting with Post-LASSO selection of subgroup–covariate interactions to achieve exact covariate balance both overall and within subgroups. Building on this, Yang et al. extended these methods to time-to-event outcomes, integrating OW with LASSO selection to enable robust subgroup analyses of survival endpoints.

**Software Implementation**

The increasing methodological developments around OW and other balancing weights have been facilitated by accessible software. Zhou et al. introduced *PSweight*, an R package that implements a comprehensive suite of propensity score weighting methods.[41] The package supports estimation of treatment effects for binary and multiple treatments, incorporates a variety of balancing weights including IPTW, overlap weights, matching weights, and entropy weights, and provides tools for both continuous and survival outcomes. Importantly, *PSweight* includes functions for variance estimation, diagnostic checks, and visualization of covariate balance, making advanced weighting methods, such as OW and their extensions, readily available to applied researchers. By lowering the barrier to implementation, this software has accelerated the adoption of OW in epidemiology, pharmacoepidemiology, and health services research. Furthermore, the recently developed R package *PSsurvival* extends propensity score weighting methods specifically to time-to-event outcomes.[29] The package accommodates binary and multiple treatment settings and provides tools for estimating both causal survival curves and marginal hazard ratios via weighted survival models, including weighted marginal Cox regression.

**Combining OW with Propensity Score Stratification**

Thomas et al. examined the relationship between propensity score stratification (PSS) and OW in the presence of substantial covariate imbalance.[42] They showed that OW generally provides superior bias–variance performance compared with IPTW and standard stratification, particularly when overlap is limited. Importantly, they demonstrated that a Mantel–Haenszel form of PSS can be viewed as a coarsened version of OW, yielding nearly equivalent performance in many settings.

**OW under Propensity Score Model Misspecification**

In practice, true propensity scores are unknown in non-randomized studies and must be estimated, raising the risk of model misspecification. Zhou et al. conducted extensive simulation studies to evaluate the performance of propensity score methods under conditions of limited overlap and propensity score model misspecification, comparing IPTW with OW, matching weights, and entropy weights.[43] They found that OW, matching weights, and entropy weights consistently outperformed IPTW across all metrics, yielding lower bias, smaller root mean squared error, and better confidence interval coverage. Notably, although OW and similar approaches target a different estimand (the overlap population) they demonstrated remarkable robustness to model misspecification.

**Dynamic Treatment Regimes**

Dynamic treatment regimes (DTRs) provide a framework for estimating optimal, individualized treatment strategies that adapt to a patient's evolving characteristics over time, representing a cornerstone of precision medicine. One important methodological advance in this area is doubly-robust weighted least squares (dWOLS), first introduced by Wallace and Moodie.[18] Although not originally framed in those terms, dWOLS in fact employs OW scheme, combining outcome regression with OW to yield doubly robust estimators that are consistent if either the treatment or outcome model is correctly specified. This framework has since been extended to increasingly complex settings: Schulz and Moodie applied dWOLS to determine optimal dosing strategies (with generalized OW scheme)[44], while Simoneau et al. adapted it for time-to-event outcomes.[45]

**Translational Application of OW**

Several translational papers have made substantial contributions and highlighted the practical relevance of OW and related propensity score methods for clinical and epidemiologic research. Thomas et al. described OW as a method that mimics key features of randomized trials by emphasizing patients at clinical equipoise, thereby producing target populations that are both interpretable and policy-relevant.[8] In a companion piece, they outlined how different propensity score methods, including OW, can be used more broadly to define different target populations in observational research, guiding investigators to match analytic choices with clinical questions.[46] Desai and Franklin provided a practitioner-oriented primer comparing IPTW, OW, and other weighting schemes, and outline key considerations for selecting and implementing an appropriate propensity score weighting approach for confounding adjustment.[47] Extending this line, Li et al. discussed the role of estimands in observational studies, building on the ICH E9 (R1) framework, and outlined practical considerations for choosing between ATE, ATT, and ATO estimands.[48] More recently, Rizk et al. provided a practical guidance on when overlap weighting is appropriate in real-world studies, emphasizing that its use should be aligned with the research question, target estimand, and target population of interest.[49] Collectively, these contributions highlight the translational importance of overlap weighting (OW): making advanced statistical methods more accessible, linking them to trial-like interpretations, and situating them within contemporary regulatory and methodological guidance.

An important and sometimes underappreciated feature of OW is that the resulting target population is not fully opaque but can be directly examined.[8,50] In practice, investigators can inspect and report weighted baseline covariate tables to characterize the overlap population and assess its clinical relevance, much as one would describe the study population in a randomized trial. Making the OW population explicit through such summaries often alleviates common concerns about interpretability, and highlights that OW targets a well-defined, data-driven population concentrated in regions of clinical equipoise.

**OW in Health Equity and Racial Disparities Research**

OW has also proven useful beyond causal inference for descriptive comparisons, including the study of racial and ethnic disparities. In particular, Li et al. demonstrated that propensity score methods, including OW, can be used to compare groups on measured covariates even when the assumption of unconfoundedness does not hold, as long as the estimand and interpretation are made explicit.[51] Specifically, OW reweights the sample to a target population defined by the region of covariate overlap, allowing fairer descriptive comparisons across racial or ethnic groups while minimizing extrapolation to non-overlapping covariate regions.

**WHEN TO USE OVERLAP WEIGHTING**

OW has rapidly evolved from a theoretical modification to propensity score weighting into a general design principle that enhances causal inference in both observational studies and randomized trials. The defining property of OW is its focus on individuals at empirical equipoise (i.e., those who could plausibly receive either treatment), thereby yielding estimates that are stable, efficient, and interpretable, in the presence of positivity violations.

First, a key strength of OW is that it adapts the target estimand to the empirical support of the data. The resulting data-adapted OW population can be characterized using a weighted baseline covariate table, constructed prior to seeing any outcome data.[50] Consistent with the principle of separating design from analysis, this characterization constitutes a design step. Accordingly, the decision to use or not use OW can be informed by the clinical relevance of the observed OW population, thereby grounding the choice of estimand in the data rather than in a purely hypothetical target. When covariate overlap is strong, OW converges to the ATE and offers variance gains over IPTW. As overlap deteriorates, OW continues to produce stable estimates by targeting the ATO. Under treatment effect heterogeneity, the efficiency gains of OW relative to IPTW are largest when overlap is weakest, precisely where IPTW becomes unstable and sensitive to modeling choices. Thus, OW operates as both a precision-enhancing method under good overlap and a principled design strategy under poor overlap. Figure 1 provides a conceptual illustration of how estimand alignment and efficiency gains evolve with the degree of covariate overlap.

Second, OW offers demonstrable advantages for treatment-effect discovery, particularly in exploratory analyses or when the goal is to detect whether a treatment confers benefit in a clinically meaningful subgroup. As shown by Mao et al.[33], OW yields higher power than IPTW and other weighting strategies in the presence of limited overlap, while maintaining valid uncertainty quantification through appropriate variance adjustment and augmentation. In early-phase comparative effectiveness studies and exploratory real-world data analyses, OW therefore emerges as a principled approach to maximize sensitivity to treatment signals without sacrificing inferential validity.

Third, OW naturally mimics randomized trial design logic by emphasizing units at clinical equipoise as well as achieving improved balance in covariates. This aligns with clinical trial design principles and the modern target trial emulation framework[9]. In both randomized trials with chance imbalance and observational studies designed to emulate target trials, OW recovers the treatment effect for a trial-eligible population (compared to other weighting schemes), those for whom the treatment choice is at least partially uncertain, thereby producing clinically interpretable estimands grounded in real-world treatment decision contexts.

Fourth, OW provides a transparent, principled alternative to trimming and truncation, which are pervasive in applied work yet often implemented in an *ad-hoc* manner. Trimming discards data and implicitly shifts the estimand, while different thresholds can lead to different conclusions. OW instead smoothly down-weights observations with extreme propensity scores, preserving all data, defining an interpretable estimand, and delivering the stability and robustness that trimming seeks without arbitrary tuning. Thus, OW can be favored since it avoids subjective decisions and improves reproducibility.

These considerations suggest that OW is not merely a remedy for non-overlap but an estimand-transparent strategy for efficient, design-driven causal inference across study settings. Even when IPTW remains the primary analytic method, OW should be used as a secondary analysis: meaningful differences between ATE and ATO estimates signal potential positivity violations or effect heterogeneity and should prompt reflection on the choice of target population, target estimand, and study design.

**Table 1: Summary of Existing Developments on Overlap Weight**

| Topic | Main Contributions | Key References |
|---|---|---|
| **Formalization of OW for causal inference** | OW and the Average Treatment Effect in the Overlap Population (ATO) were formally introduced within the *balancing weights* framework, unifying IPTW, ATT, and matching weights. OW minimizes asymptotic variance and achieves exact mean balance under logistic PS. | Li et al.[6] |
| **Simulation Evidence** | Systematic evaluations show OW outperforms IPTW with/without trimming in bias–variance trade-off and balance. Austin and Liu extended inference and power calculation methods for OW, providing closed-form and bootstrap variance estimators. | Li et al.[7]; Austin[19,20]; Liu et al.[21] |
| **OW with Multiple Treatments** | Generalized OW extend balancing weights to multi-arm settings, emphasizing units likely to receive any treatment and minimizing total variance for pairwise contrasts. | Li & Li[22]; Yu et al.[24] |
| **OW with Censored Time-to-Event Outcomes** | OW incorporated into survival analyses via IPCW and pseudo-observation methods, improving efficiency in estimating survival and restricted mean survival time. | Mao et al.[25]; Cheng et al.[30]; Cao et al.[31]; Zeng et al.[32] |
| **Combining OW with Outcome Regression** | (1) Augmented OW integrates outcome models into OW for double robustness, improving precision and consistency under misspecification. | Mao et al.[33]; Zeng et al.[32] |
| | (2) Regression of (Y) on treatment and PS ((e(x))) yields an ATO-estimating coefficient. Vansteelandt & Daniel formalized this equivalence, and Ding later unified OW with PS-adjusted OLS regression. | Vansteelandt & Daniel[17]; Ding[34] |
| **OW for Randomized Clinical Trials** | In RCTs, OW improves finite-sample precision despite asymptotic equivalence with IPTW; extended to subgroup, win statistic, and cluster RCT analyses. | Zeng et al.[35]; Yang et al.[36] |
| **OW in Causal Subgroup Analysis** | Introduced subgroup-weighted ATE (S-WATE) and OW-pLASSO algorithm, balancing overall and within-subgroup covariates; extended to time-to-event outcomes. | Yang et al.[36,39,40] |
| **Software Implementation** | *PSweight* R package implements OW and other balancing weights with diagnostics and visualization tools, promoting accessibility for applied researchers. *PSsurvival* R package extends the application to time-to-event outcomes | Zhou et al.[41]; Yang et al.[29] |
| **Combining OW with Stratification** | Demonstrated that Mantel–Haenszel-type PS stratification is a coarsened version of OW with similar performance and better interpretability. | Thomas et al.[42] |

| **OW under PS Model Misspecification** | Simulations reveal OW, matching, and entropy weights outperform IPTW under PS misspecification—showing OW's robustness and stability. | Zhou et al.[43] |
|---|---|---|
| **Translational Application of OW** | Translational papers emphasized OW's trial-like interpretation, practical relevance, and role in defining estimands aligned with research questions and ICH E9 (R1) guidance; Rizk et al. further provided practical guidance on aligning OW use with study aims and target populations. | Thomas et al.[8,46]; Desai & Franklin[47]; Li et al.[48]; Rizk et al.[49] |
| **Dynamic Treatment Regimes (DTRs)** | dWOLS extends OW to dynamic treatment regimes, combining outcome regression and OW for double robustness and application to dosing and survival outcomes. | Wallace & Moodie[18]; Schulz & Moodie[44]; Simoneau et al.[45] |
| **OW in Racial Disparities Research** | OW enables descriptive comparisons across racial and ethnic groups without assuming unconfoundedness, focusing on populations with overlapping covariate distributions. | Li & Li.[51] |

**Figure Captions**

**Figure 1.** Illustrations of perfect, good, and weak overlap in propensity score distributions between treatment and control groups

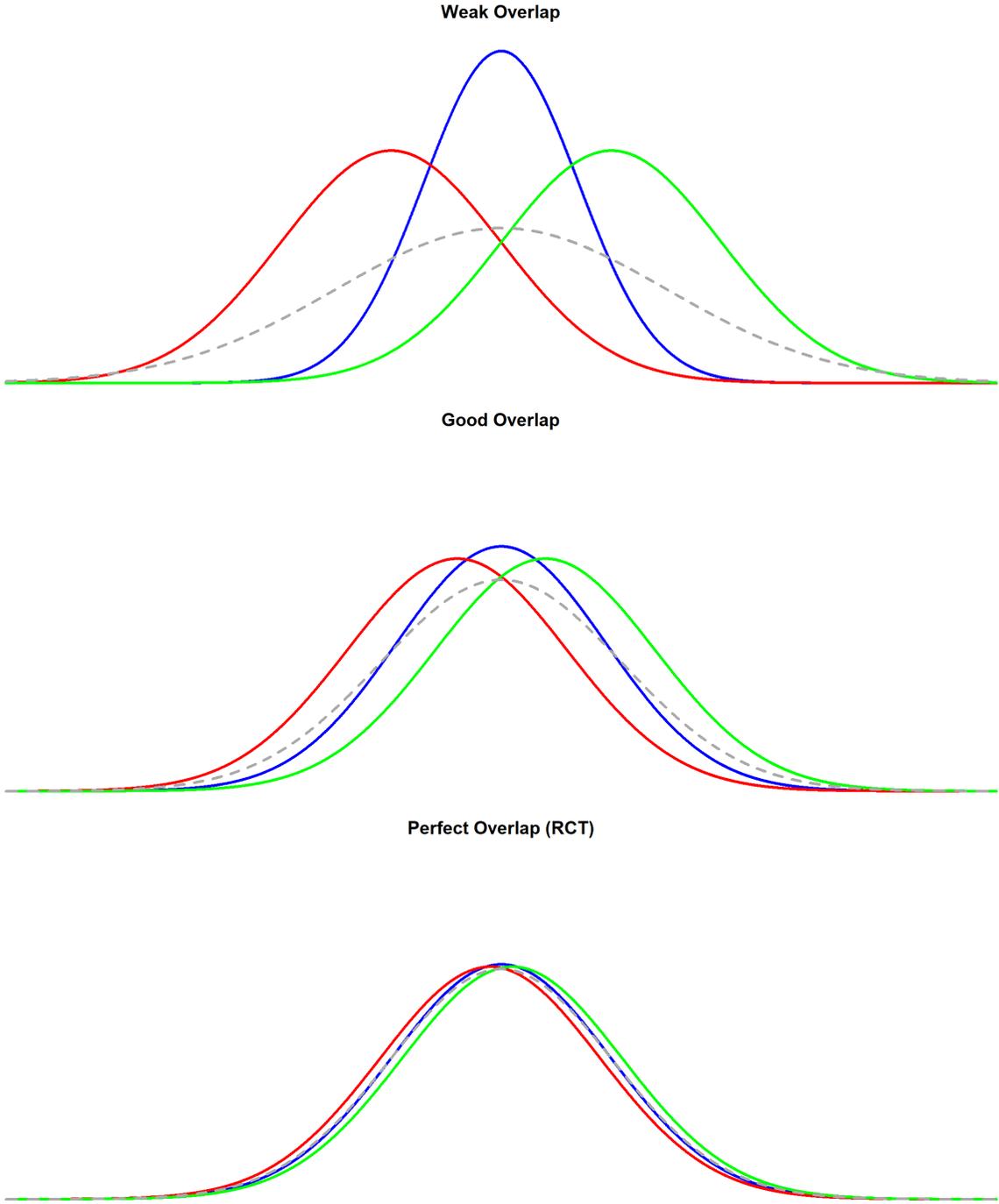